\documentclass[journal=nalefd,manuscript=letter]{achemso}

\usepackage[version=3]{mhchem} 
\usepackage{textcomp}
\usepackage{wrapfig}

\newcommand{\onlinecite}[1]{\hspace{-1 ex} \nocite{#1}\citenum{#1}}
\author{Klaus~D.~J\"ons}
\affiliation[IHFG, University of Stuttgart]
{Institut f\"ur Halbleiteroptik und Funktionelle Grenzfl\"achen, University of Stuttgart, Allmandring 3, 70569 Stuttgart, Germany}
\alsoaffiliation[current address: Kavli Institute of Nanoscience, Delft University of Technology, 2600 GA Delft, The Netherlands]{Research Center SCoPE}
\altaffiliation{These authors contributed equally to this work}
\author{Ulrich~Rengstl}
\altaffiliation{These authors contributed equally to this work}
\email{u.rengstl@ihfg.uni-stuttgart.de}
\author{Markus~Oster}
\author{Fabian~Hargart}
\author{Matthias~Heldmaier}
\author{Samir~Bounouar}
\author{Sven~M.~Ulrich}
\author{Michael~Jetter}
\author{Peter~Michler}
\affiliation[IHFG, University of Stuttgart]
{Institut f\"ur Halbleiteroptik und Funktionelle Grenzfl\"achen, University of Stuttgart, Allmandring 3, 70569 Stuttgart, Germany}

\title[Monolithic on-chip integration of semiconductor waveguides, beamsplitters and single-photon sources]
  {Monolithic on-chip integration of semiconductor waveguides, beamsplitters and single-photon sources}

\abbreviations{QD, MOVPE, SEM, DBR, HSQ, ICP, RIE, FDTD}
\keywords{quantum dots, single photons, waveguides, photonic integrated circuits, integrated beamsplitter}

\begin{document}


\begin{abstract}
The implementation of a fully integrated Hadamard gate on one single chip is currently one of the major goals in the quantum computation and communication community. Prerequisites for such a chip are the integration of single-photon sources and detectors into waveguide structures such as photonic crystals or slab and ridge waveguide.
Here, we present an implementation of a single-photon on-chip experiment based on a III-V semiconductor platform. Individual semiconductor quantum dots were used as pulsed single-photon sources integrated in ridge waveguides, and on-chip waveguide-beamsplitter operation is verified on the single-photon level by performing off-chip photon cross-correlation measurements between the two output ports of the beamsplitter. A careful characterization of the waveguide propagation losses ($\sim 0.0068$\,dB/\textmu{}m) documents the applicability of such GaAs-based waveguide structures in more complex photonic integrated circuits. The presented work marks an important step towards the realization of fully integrated photonic quantum circuits including on-demand single-photon sources.
\end{abstract}

\section{Introduction}
Shortly after the pioneering work of Knill, Laflamme and Milburn which opened the field of linear optics quantum computation~\cite{Knill.Laflamme.ea:2001}, first optical two-photon gates have been realized~\cite{O'Brien.Pryde.ea:2003,Gasparoni.Pan.ea:2004}. 
These experimental demonstrations of such an optical controlled-NOT gate have been made with bulky optics. However, the applicability of quantum information science~\cite{Bennett.DiVincenzo:2000} relies strongly on the on-chip integration, i.e. miniaturization, of such devices. 
In 2008, the first on-chip quantum logic gate has been reported by Politi et al.,~\cite{Politi.Cryan.ea:2008} and several important applications followed,\cite{Politi.Matthews.ea:2009,Peruzzo.Shadbolt.ea:2012,Spring.Metcalf.ea:2013,Broome.Fedrizzi.ea:2013} showing the power of integrated photonic quantum circuits. 
However, all these previous studies have been realized by use of external photon sources based on probabilistic parametric down conversion.\cite{Politi.Cryan.ea:2008}
On the basis of silicon substrates, an integration of on-demand single-photon sources such as single quantum emitters, is exceptional rare\cite{Benyoucef.Lee.ea:2009,Wiesner.Schulz.ea:2012} yet, whereas waveguide integration has not been reported to the best of our knowledge.
One approach to integrate the photon source on chip is integrated waveguide four-wave mixing.~\cite{Sharping.Lee.ea:2006,Silverstone.Bonneau.ea:2014} However, these sources rely on an interaction length, which is very large in terms desirable of on-chip scales and they deliver Poissonian photon statistics.
Nonlinearities can also be used with III-V-semiconductor-based photon sources, for example by integrated spontaneous parametric down-conversion in GaAs/AlGaAs Bragg reflection waveguides.~\cite{Horn.Abolghasem.ea:2012}

On the other hand, III-V-semiconductors are also perfectly suitable for the integration of quantum dots (QDs) as triggered single-photon\cite{Michler.Kiraz.ea:2000} and entangled photon pair sources.~\cite{Young.Stevenson.ea:2006,Akopian.Lindner.ea:2006}
Using GaAs as the platform, very efficient waveguiding in photonic crystal (PC) structures has been shown.\cite{Sugimoto.Tanaka.ea:2004} The coupling of single QDs to the propagation mode of the PC waveguide additionally enhances their spontaneous emission rate.\cite{Lund-Hansen.Stobbe.ea:2008} Indeed, very efficient guiding of their single-photon emission has been shown\cite{Schwagmann.Kalliakos.ea:2011}, even for the conditions of only weak Purcell enhancement\cite{Laucht.Putz.ea:2012}. It has been shown, that the coupling of single QDs to the waveguide mode can be enhanced by designing the PC waveguide to act as a Fabry-P\'erot cavity\cite{Laucht.Putz.ea:2012} or by coupling QDs to PC nanocavities.~\cite{Englund.Majumdar.ea:2010}

Ridge and rib-type photonic channels have also been studied on this material system, and low-loss waveguiding was reported.\cite{Ferguson.Kuver.ea:2006} Even a spin-photon interface was realized by coupling QDs to two orthogonal waveguides.\cite{Luxmoore.Wasley.ea:2013}
Additionally, it has been shown that ridge waveguides are suited for integrating superconducting single-photon detectors~\cite{Sprengers.Gaggero.ea:2011}. Using such detectors, on-chip time-resolved photon counting from an ensemble of integrated QDs has been recently reported~\cite{Reithmaier.Lichtmannecker.ea:2013}.
However, one major prerequisite for the realization of quantum photonic integrated circuits based on QDs is missing up to now, i.e. the on-chip generation of single photons from individual QDs together with the functionality of beamsplitter operations on the single-photon level.
In this letter, we present such a device using the coupling between two ridge waveguides to create an on-chip 50/50 beamsplitter. The usage of a GaAs-based heterostructure enables the monolithic integration of InGaAs/GaAs QDs as single-photon sources into the waveguides, paving the way for a fully integrated Hadamard gate.

\section{Results}
\subsection{Design of the waveguide structure}

\begin{figure}[t!]
	\begin{center}
		\includegraphics[width=0.9\textwidth]{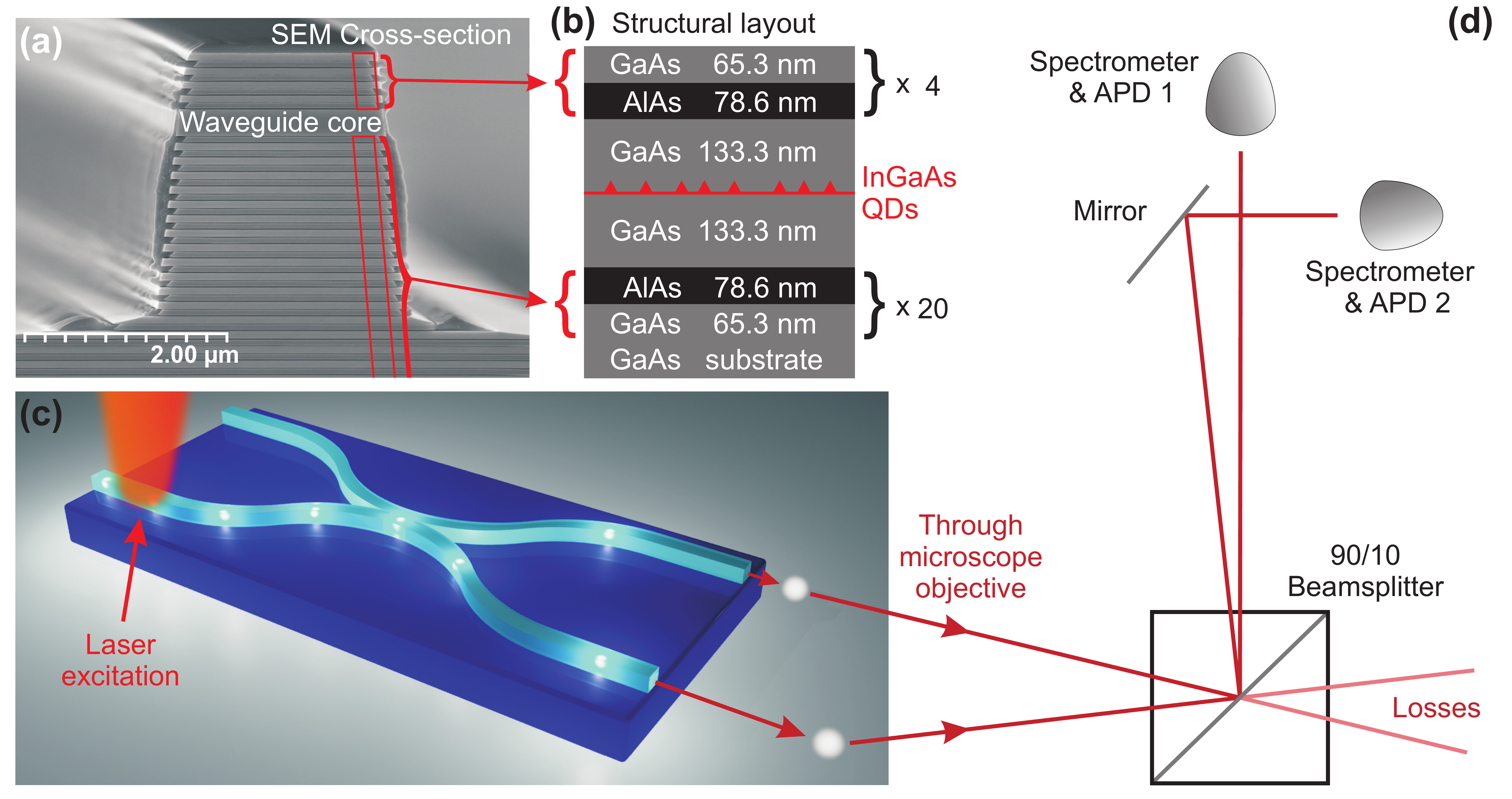}
	\end{center}
	\caption{a) Scanning electron microscope (SEM) image of the cleaved facet of a Bragg reflection waveguide structure with 4 DBR pairs on top and 20 DBR pairs at the bottom of the waveguide core. b) Schematical layer structure. c) Visualization of the integrated 50/50~beamsplitter with a directional, multi-mode coupler in the middle, surrounded by cosine-shaped bends. The embedded QDs are excited in only one arm, wheras the emission is detected from both arms. d) Diagram of the optical setup. Both output ports of the waveguide structure can be mapped to different spectrometers.}
	\label{Fig:WaveguideStructure}
\end{figure}

The layer structure of the presented device was grown by metal-organic vapor-phase\linebreak epitaxy~(MOVPE) using a (100)-GaAs substrate, with a miscut of 6$^\circ$ to the (111)A facet. Figure\,\ref{Fig:WaveguideStructure}(a) shows a scanning electron microscope~(SEM) image of the cross-section of the finalized structure, where the grown layer sequence (see schematic drawing in Fig.\,\ref{Fig:WaveguideStructure}(b)) is clearly visible. It consists of 20 pairs of distributed Bragg reflectors~(DBR) at the bottom, which are obtained by the alternating deposition of AlAs and GaAs layers. On top of this DBR structure the waveguide core is deposited, consisting of a 267\,nm thick GaAs layer. The usage of III-V semiconductor material enables the implementation of InGaAs/GaAs QDs as single-photon sources using the self-organizing Stranski-Krastanow growth mode. To enhance the coupling efficiency between QDs and the waveguide, they were grown in the center of the GaAs core layer. The vertical confinement is finished by an additional set of 4 pairs of DBRs at the top. The DBR structure was optimized for vertical detection of the QD emission around 930\,nm, which allowed detailed pre-characterization.~\cite{Richter.Hafenbrak.ea:2010, Ulhaq.Weiler.ea:2012}

A sketch of the used waveguide structure is shown in Fig.\,\ref{Fig:WaveguideStructure}(c). In contrast to Ref.~\onlinecite{Politi.Cryan.ea:2008}, we are using multi-mode waveguides with a width of 2\,\textmu m and a directional multi-mode coupler without a gap between the waveguides. 
The waveguide design was optimized using a commercial beam propagation software (RSoft Photonics Suite). The coupler has a nominal length of 118.5\,\textmu m, which should lead to a 50/50 splitting ratio at 910\,nm. This coupler is surrounded by cosine shaped bends to separate the two beams to a distance of 50\,\textmu m within 437.1\,\textmu m. Due to the dependency of the splitting ratio on the coupler length, the merging regions of the waveguides have to be created with high accuracy.

Therefore, the structure was written using electron beam lithography with a negative tone resist based on hydrogen silsesquioxane~(HSQ), which forms a highly resistive SiO$_2$ mask after developing. The subsequent structuring by an inductively coupled plasma reactive ion etching (ICP-RIE) ICP-100 system (Oxford Instruments) was carried out using chlorine-based chemistry. This step was monitored via an in-situ laser interferometer which allowed the precise stop of etching in the GaAs layer of the 5th mirror pair of the bottom DBR. A cleaved facet of the etched waveguide is shown in Fig.\,\ref{Fig:WaveguideStructure}(a). The obtained sidewalls have an angle of less then 7$^\circ$ to the vertical, which is introduced by mask erosion during the etching process. 

\subsection{Optical characterization and comparison with theory}

\begin{figure}[t]
	\begin{center}
		\includegraphics[width=0.45\textwidth]{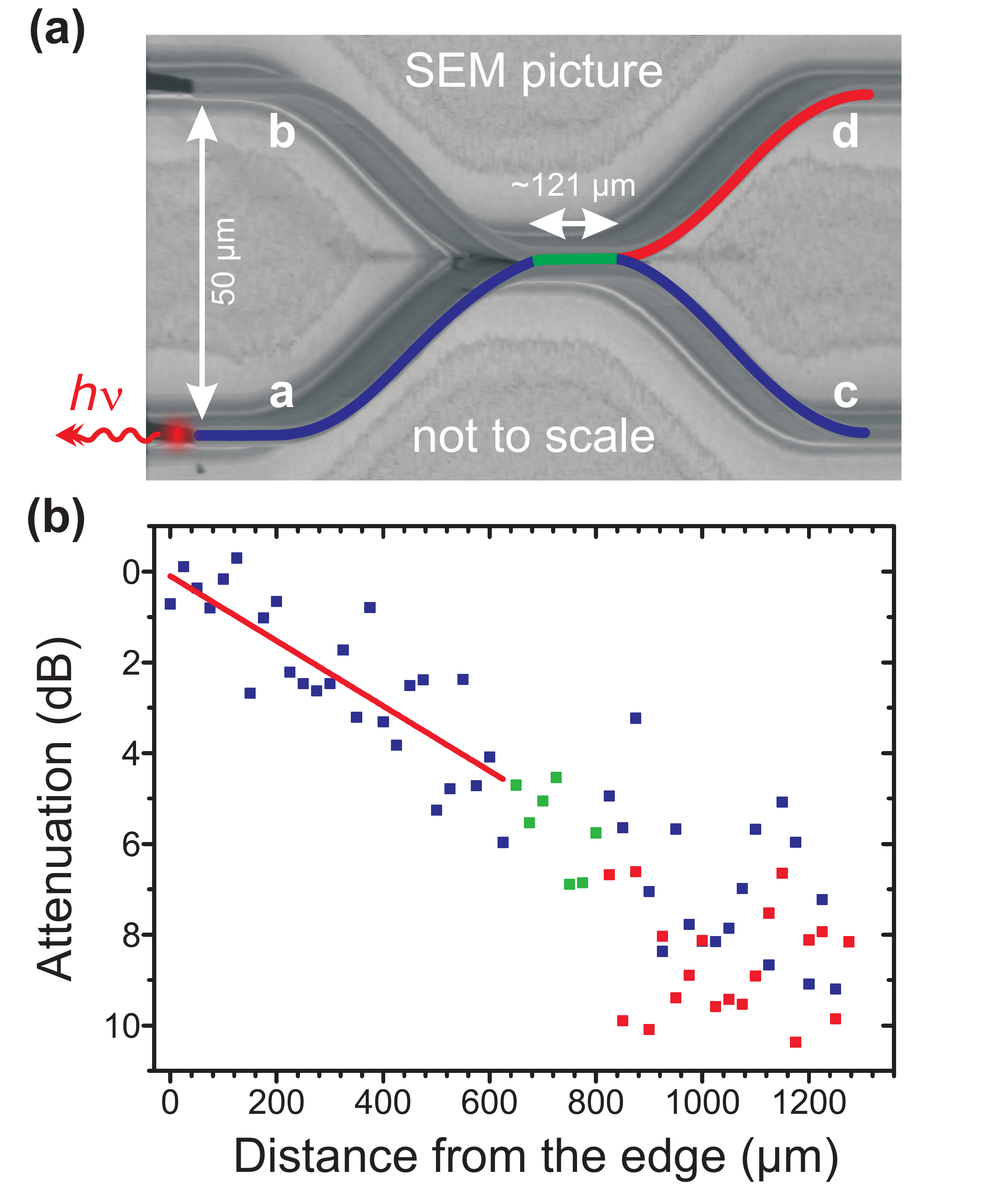}
	\end{center}
	\caption{a) SEM picture of a measured waveguide. For better visualization, the image is stretched in vertical direction. b) Attenuation of the QD emission in dependence on the distance between the QD and the cleaved facet of detection arm "a". The corresponding beam spot position is marked in the SEM picture (a) (Blue:~QD-excitation on arm "a" and "c", red:~excitation on arm "d", green:~excitation in the coupler region).}
	\label{fig:Absorption}
\end{figure}

For the optical characterization, the sample was cleaved perpendicular to the waveguides, as shown in Fig.\,\ref{Fig:WaveguideStructure}(a), leaving a 200\,\textmu m long straight section before the first bended section. The sample was placed on a motorized stage inside a He-flow cryostat and cooled to 5\,K. The excitation was carried out from the top with a fiber coupled titanium-sapphire laser through a microscope objective with an NA of 0.45. The minimal spot diameter of the excitation beam was approximately 2\,\textmu m on top of the sample. The fiber coupler and microscope objective were placed on a motorized stage to allow independent optimization of laser excitation and detection. The emission from the waveguides was measured at an angle of 90$^\circ$ through a fixed microscope objective with an NA of 0.45. For cross-correlation measurements, we have split the collected photons from the individual waveguide output ports after the objective and mapped them simultaneously to two different spectrometers attached with single-photon detectors (Fig.\,\ref{Fig:WaveguideStructure}(d)).

To estimate the structural quality of the etched structure, the optical losses of the waveguide were determined. Using a continuous wave laser at 807\,nm, above the GaAs barrier, the QDs were excited to the saturation level. The beam spot was scanned along the waveguide and we measured the maximum single QD intensity through the cleaved edge of the waveguide arm "a", as shown in Fig.\,\ref{fig:Absorption}(a). These measurements were performed in three different configurations, i.e. excitation on the detection arm "a" in front of the beamsplitter, within the coupling region itself as well as on both arms "c" and "d" behind the coupler. Figure\,\ref{fig:Absorption}(b) shows the corresponding maximal measured intensity of 70~QDs. The losses of the structure are determined by an exponential fit of the measured intensities over their distances to the cleaved facet. This method is prone to fluctuations in the quality of the single QDs and their coupling efficiency into the waveguide, which leads to a high spreading of the measurement points. After the beamsplitter, the wavelength-dependent splitting ratio adds additional errors to the measurement by inhibiting a clear observation of the expected 50\,\% reduction in intensity inside the coupler region. Due to this, we have only considered measurements for excitation in arm "a" in front of the beamsplitter to determine the losses of the bare waveguide. We were able to estimate the average losses in the waveguide structure as $(0.0068\pm0.0005)$\,dB/\textmu m, which is in good agreement with recently obtained values on GaAs-based ridge waveguide structures.\cite{Reithmaier.Lichtmannecker.ea:2013}

In the following, we have estimated the on-chip efficiency of the device and used it to predict an overall efficiency for the whole setup. We define the overall efficiency as the probability to detect one photon of a certain QD through one detection arm ("c" or "d") per excitation pulse of the pump laser. However, the on-chip efficiency is the probability that a photon from this QD reaches the end of one detection arm ("c" or "d") before coupling into free space.
One major factor which limits the on-chip efficiency is the coupling of the quantum dot luminescence into the waveguide. This efficiency has been estimated by two-dimensional (range: 100\,\textmu m) and three-dimensional (range: 40\,\textmu m) finite-differential time-domain~(FDTD) simulations 
using the free software package Meep.\cite{Oskooi.Roundy.ea:2010} We obtain a theoretical coupling efficiency of a perfectly centered and aligned dipole into the waveguide core of $\beta = (7\pm1)$\,\%. For QDs with a distance of 915\,\textmu m to the cleaved facet, like the one used for the cross-correlation measurement below, we obtain an additional attenuation of $(78\pm5)$\,\% from Fig.\,\ref{fig:Absorption}(b). 
If we assume an ideal quantum efficiency of our QDs ($\eta = 100\,\%$), this would lead to an on-chip efficiency of $(1.6\pm0.6)$\,\% at each output port ("a" or "b") before emission into free space. 
However, for the overall efficiency we have to take into account this coupling efficiency from the output port ("a" or "b") into free space. This efficiency is again obtained by two-dimensional FDTD simulations and determined to be
less then $(6.8\pm1.0)$\,\%. 
These high losses originate mainly from the total internal reflection in the horizontal plane, where the structure is designed as a multimode waveguide.  
By additionally taking into account the collection efficiency of our objective of 33\,\% and the total measured efficiency of our optical setup, including the APDs, of $(6.0\pm0.5)$\,\% for detection around 910\,nm, we can derive a theoretical overall efficiency of $(0.0021\pm0.0007)$\,\%.
This efficiency might be reduced due to additional scattering effects on the cleaved facet and a possible internal quantum efficiency of the QDs below one,\cite{Johansen.Stobbe.ea:2008,Stobbe.Schlereth.ea:2010} but is already in good agreement with the measured efficiencies as shown below.

\begin{figure}[t]
	\begin{centering}
		\includegraphics[width=0.9\textwidth]{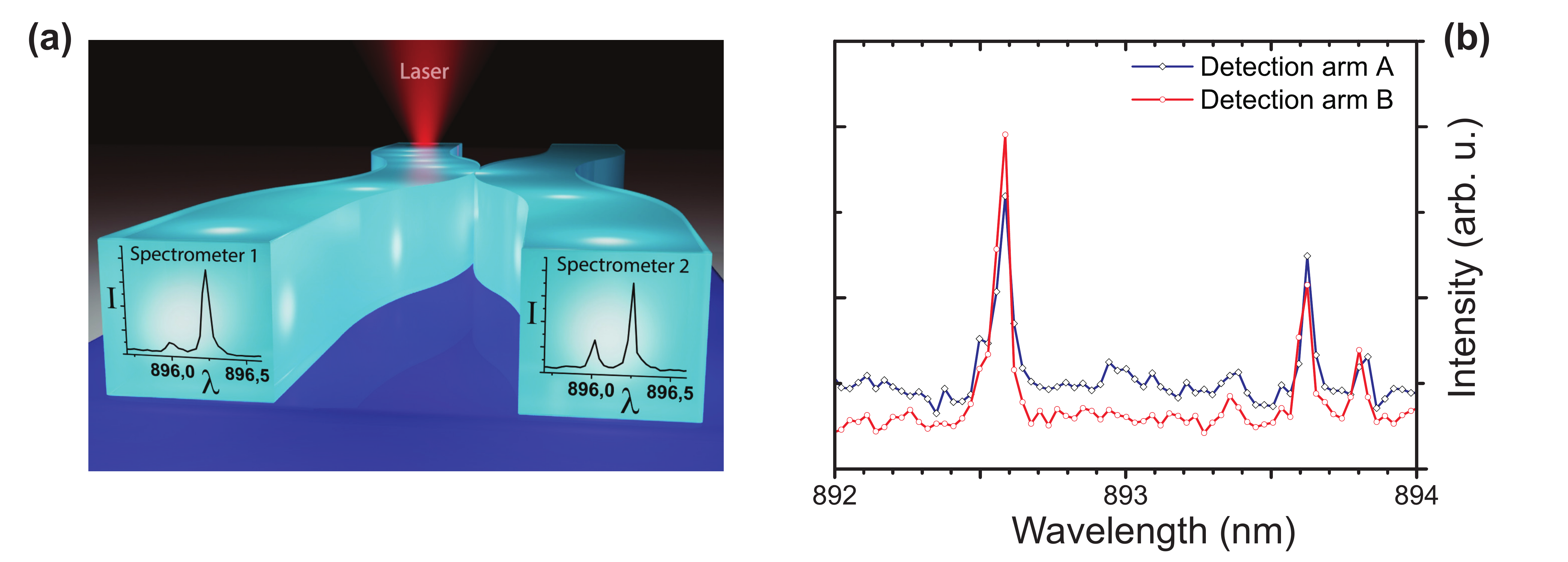}
	\end{centering}
	\caption{Section of typical spectra received simultaneous on both separate detection arms ("a" and "b"), while exciting on arm "c" behind the coupler. a) Spectra of a single quantum dot recorded with the setup explained in Fig.\,\ref{Fig:WaveguideStructure}(d) featuring two different spectrometers in the two detection paths. Since the two detections paths are not identical, a comparison of the intensities is not feasible. In graph b) both spectra were collected with the same detection path to eliminate path dependent losses. Three QD emission lines show up in both detection arms with nearly 50/50 splitting ratio. 
	}
	\label{fig:Splitting}
\end{figure}

Figure\,\ref{fig:Splitting} depicts quantum dot spectra collected from the end facets of the detection arms ("a" and "b") while exciting in arm "c" of the beamsplitter. Graphic\,\ref{fig:Splitting}(a) shows spectra collected simultaneously with two different spectrometers like explained in Fig.\,\ref{Fig:WaveguideStructure}(d). This setup configuration is used to perform cross-correlation measurements of the two output arms of the waveguide. However, since two spectrometers are used we cannot extract the beamsplitter ratio from these measurements.
Instead, the splitting ratio of the waveguide beamsplitter has been determined by measuring the signal from both detection arms ("a" and "b") using the same detection path and detector alternately on both arms "a" and "b".
The detection signal was optimized on the same QD for both measurements while keeping the excitation power constant. The excitation laser was set to pulsed operation with 2\,ps-pulses and tuned to 864\,nm for pumping via the wetting layer. Figure\,\ref{fig:Splitting}(b) shows two QD spectra taken at the two different output ports ("a" and "b"). Three emission lines with comparable intensities can be easily identified, which suggests a splitting ratio close to 0.5 for this wavelength region around 894\,nm.
This corresponds to an effective length of the coupler of $(122\pm2)$\,\textmu m, which is in good agreement with the designed coupler length of 118.5\,\textmu m and with the SEM-measurements showing a value of $(121^{+13}_{-1})$\,\textmu m. However, the wavelength dependency of the beamsplitter ratio does not follow a simple theoretical model. One has to take into account different wavelength dependencies for higher optical modes in the waveguide. It is hard to experimentally determine the emission mode of a QD, since we do not know the exact position and orientation of the QD dipole moment inside the waveguide.

\begin{figure}[t]
	\begin{centering}
		\includegraphics[width=0.90\textwidth]{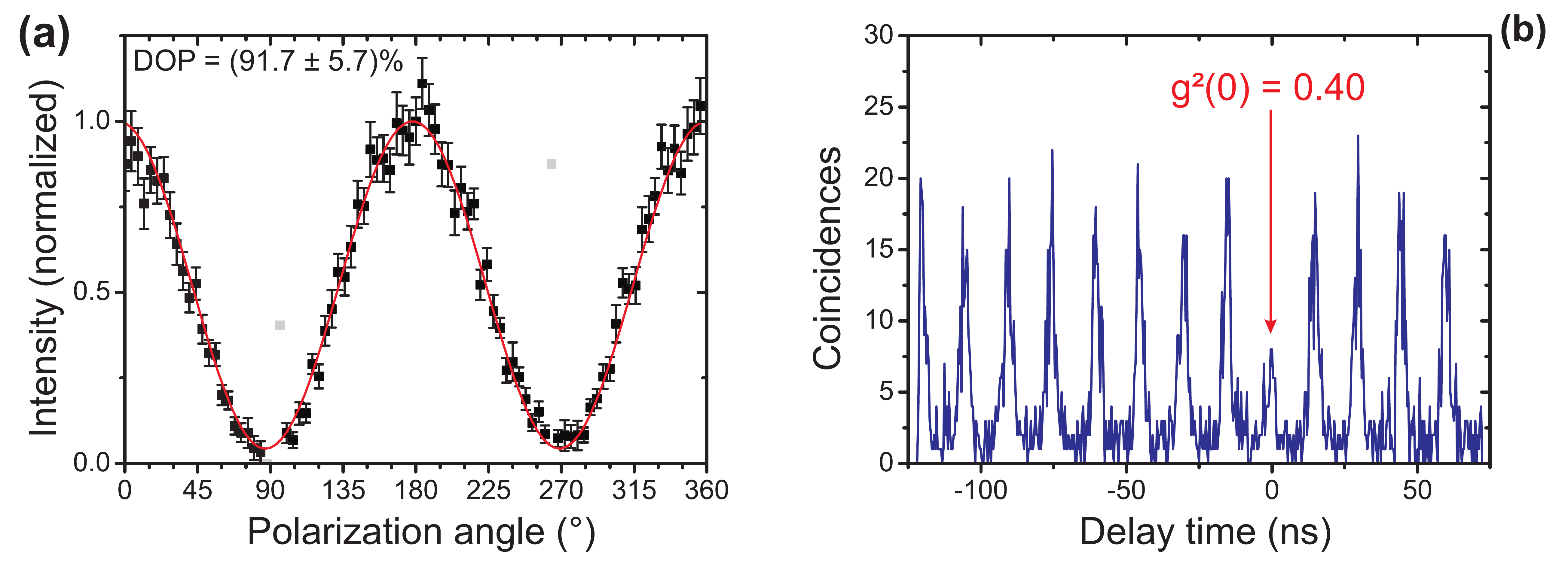}
	\end{centering}
	\caption{a) A high degree of polarization of the emission from the QDs is maintained through the waveguide. b) Raw data of a second-order cross-correlation function betwwen both output arms "a" and "b" of the waveguide during pulsed excitation of a QD behind the coupler (arm "c"). The measurement shows a clear antibunching behavior. After background correction a g$^{(2)}(0)$ value of 0.4 is obtained.
}
	\label{fig:Polarization-g2}
\end{figure}

Polarization-dependent measurements were performed during continuous wave excitation via the wetting layer by using a $\lambda$/2-waveplate and a Glan-Thompson polarizer. Most QDs show a large degree of polarization (DOP) above 90\,\% (Fig.\,\ref{fig:Polarization-g2}(a)), which corresponds to the expected behavior assuming a heavy-hole exciton dipole moment orientated parallel to the sample surface~\cite{Bester.Nair.ea:2003,Huo.Witek.ea:2014}. The non-unity DOP which is found over the whole waveguide can be explained by two possible mechanisms. 
One possibility is the ground state mixing between light holes and the predominant heavy holes. This mixing could be a result of the partly oxidized, strained AlAs layers of the DBR structures. A contribution of light-hole excitons in the emission would lead to an additional bright state with the dipole moment along the growth direction~\cite{Bester.Nair.ea:2003,Huo.Witek.ea:2014,Plumhof.Trotta.ea:2013}. The other mechanism is the polarization-dependent loss of the guided light inside the waveguide together with polarization rotation due to sloped and rough sidewalls.\cite{Rahman.Somasiri.ea:2002,Morichetti.Canciamilla.ea:2010} This mechanism will be enhanced in the curved sections of the beamsplitter.\cite{Obayya.Rahman.ea:2001}
Both processes lead to the creation of a non-zero polarization in growth direction, thus decreasing the DOP.

To verify the beamsplitter operation of our device on a single-photon level, we have also performed cross-correlation measurements between the two output ports ("a" and "b") of the waveguide, using the photoluminescence signal of a single QD located in arm "c" behind the coupler region.
The single photons emitted by the QD are split on-chip and guided to the two output ports of the waveguide.
The photons from each port are separately send to two spectrometers and detected with single-photon detectors (Fig.\,\ref{Fig:WaveguideStructure}(d)). The excitation was again performed via the wetting layer with a ps-pulsed laser (66\,MHz repetition rate) and an average power of 10.6\,\textmu W. We have obtained count rates of 700\,counts/s and 1000\,counts/s on the avalanche photo diodes (APDs) with dark count rates of 50\,counts/s and 60\,counts/s, respectively.
These  count rates corresponds to a measured overall efficiency of $1.0\cdot10^{-5}$ to $1.4\cdot10^{-5}$ for the whole setup, which is in good agreement with the theoretical obtained value of $(2.1\pm0.7)\cdot10^{-5}$.
The cross-correlation between both APDs is shown in Fig.\,\ref{fig:Polarization-g2}(d). After subtraction of the accidental coincidences due to the intrinsic dark counts of the used avalanche photo diodes, we obtain a g$^{(2)}(0)$ value of 0.40, which clearly exhibits the preserved single-photon characteristics of a single QD. This verifies the functionality of our beamsplitter on a single-photon level.

\subsection{Discussion}
In summary, we have demonstrated a monolithically fabricated beamsplitter operating on single photons generated by QDs as integrated on-chip single-photon emitters.
The used GaAs material system is demonstrated to by highly suitable as an integration platform, showing propagation losses of only $\sim 0.0068$\,dB/\textmu{}m.
The overall efficiency of the device is in good agreement with the expected values obtained by FDTD simulations. These include a theoretical coupling efficiency of $\beta = (7\pm1)$\,\% for a perfectly aligned QD into the ridge waveguide modes. This can be enhanced in the future by the usage of photonic crystal waveguides, where coupling efficiencies of 89\,\% have been demonstrated.\cite{Lund-Hansen.Stobbe.ea:2008} Such structures suffer from higher propagation losses due to multiple scattering at fabrication imperfections. However, as has been recently outlined by P. Lohdal et al.\cite{arxiv:Lodahl.Mahmoodian.Stobbe:2014}, one solution might be a hybrid structure where the photonic crystal, in which the QDs are embedded, are coupled to ridge waveguides, where the emitted photons can propagate over longer distances.

For our device the correct beamsplitter operation on a single-photon level has be verified by cross-correlation measurements between the two output arms of the device, revealing a\linebreak\mbox{g$^{(2)}(0) < 0.5$}.
The here presented combination of the gate operation and the single-photon generation on one single III-V semiconductor chip is a further cornerstone for future quantum photonic integrated circuit applicability. 

\begin{acknowledgement}
The authors thank H.~Niederbracht and M.~Paul for advice and fruitful discussions and\linebreak T.~Schwarzb\"ack for graphical artwork. The authors acknowledge T.~Reindl in the group of J.~Weis from the MPI for solid state research Stuttgart for electron beam lithography and D.~Richter for the excellent sample growth.\\
\end{acknowledgement}

%
%

\bibliography{bibliography}

\end{document}